# Cross-View of Testing Techniques Toward Improving Web-Based Application Testing

**Mostafa Kandil[1], Ehab Hassanein[2], Sherif Mazen[3]**

[1] Information Systems department
Faculty of Computers & Information
Cairo University, Egypt
*Mostafa_kandil2003@yahoo.com*

[2] Information Systems department
Faculty of Computers & Information
Cairo University, Egypt
*e.ezat@fci-cu.edu.eg*

[3] Information Systems department
Faculty of Computers & Information
Cairo University, Egypt
*s.mazen@fci-cu.edu.eg*

**Abstract**
Web Applications (WA's) failures may lead to collapse of the institutions, therefore the importance of good quality WA's is increasing over the time. Testing is one of the best quality metrics that decide whether WA's are reliable or not. WA's testing approaches suffer from the lack of proper coverage of WA's functional requirements testing. On the other hand some approaches produce test cases that already cover WA's testing but they also produce a great number of irrelevant test cases. This research analyzed the main testing approaches for WA's and GUI applications. Also we have an overview of Test-Driven Development and its effects on the current development. The specification of good testing approach that satisfies the proper testing is then presented.

## 1. Introduction

Demand for high-quality Web Applications (WAs) continues to escalate as reliance on Web-based software increases and Web systems become increasingly complex[1]. WAs are not just these sites on the web which can get numerous number of hits by many users simultaneously, But now the trend of information systems like ERP or Management information systems (MIS) to be developed as WAs which means more complex, more sophisticated, more business rules and in big organizations over 1000 of users can use that application simultaneously. As well there are many important modules in the web such as e-payment or banking that requires very effective and efficient testing.

Since testing typically consumes 40~50% of development efforts, and consumes more effort for systems which require higher levels of reliability, it is a significant part of the software engineering[2].

In this paper we will explore some of WAs testing strategies and others of GUI testing strategies and its defined drawbacks. Also we will talk about Test-Driven Development and Test Coverage techniques. Thus we can develop new WAs testing approach.

Finally we will discuss the problems that we may find in each technique, and our future work in that area.

## 2. Testing Types

System Testing is the process to inspect and detect software systems errors and faults that restrict software required functionalities and objectives [3].

Authors in paper [3] have explored all software testing strategies in a single hierarchy diagram. Manual testing means the test of analysis documents, design documents, or algorithms before software not actually developed or in early phase of development life cycle, which it maybe Walkthrough, Review, or Inspection. Dynamic testing means the test of software code.

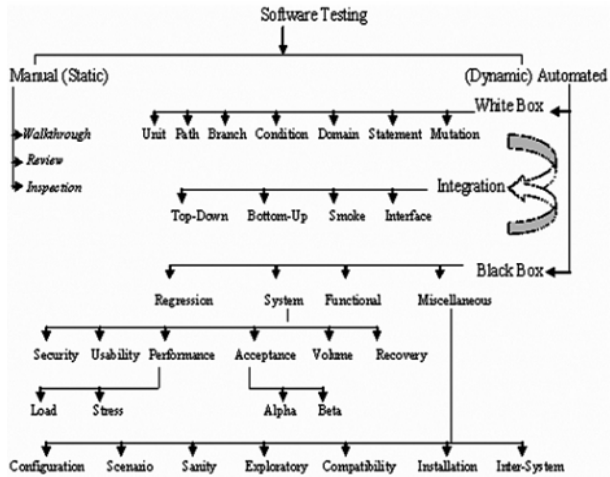

**Fig. 1 Testing Strategies in Brief [3]**

Dynamic test has three types white box testing, black box testing, and integration testing, white box testing means test the internal structure of the code, it may be unit, path, branch, condition, domain, statement, mutation.

Black box testing means testing the functionality of the application ignoring the inspection of internal code and focuses on the outputs generated in response to inputs and execution conditions. It may be functional, non-functional, system, or regression test.

Gray box or Integration testing is a combination between black-box and white-box testing, it uses systematic combination and execution of product components to insure that the interfaces between the components are correct and the product components combined are executing the software's functionalities correctly.

Automated vs. Manual testing was discussed in [3-5] the main conclusion is when using automated test it must be completed using manual test, and automatic test being less time consuming and less cost than manual. But we will ignore types of testing with respect of being automated or manual; our main concern is the technique.

## 3. Web Applications Testing

Due to the heterogeneous nature and different quality criteria of Web Applications, its components and user expectations, new demands emerge for testing of those systems to ensure a high reliability level [6]. The most important aspect of WA's is that it deals with large numbers of users, clients and stakeholders around the world. Many WA's require high quality such as banking systems, governmental, ticketing system, e-commerce etc...

### 3.1 Differences between WA's and traditional Applications

WA's differ greatly from traditional applications, as (1) WAs may consist of both dynamic and static pages which require a lot of interactions between users and web browser (2) WAs are installed across a network and they can reveal unusual flows of control (3) heterogeneous and 'dynamic' nature of the Web applications (4) users can reach any page by enter its path directly in any web browser that may affect WAs (5) there's no specific one GUI to the user (interface on Firefox maybe different than IE), because of these differences traditional testing methodologies do not apply directly to WAs. The specific features of WAs that didn't include in traditional applications must be considered to comprehend these differences in testing. Also further research efforts should be spent to define and assess the effectiveness of testing models, methods, techniques and tools that combine traditional testing approaches with new and specific ones [1, 7-10].

### 3.2 Some of WAs Testing Approaches

The first approach for WA testing was introduced in [6]. It introduced a novel approach for systematic test case generation for functional testing of WA's using a structured event-based model. Commercial test tool was used to support that approach in order to avoid costly and error-prone manual test. This approach used System-Under-Test (SUT) and its GUI to test system functionalities.

The second approach introduced by Sarah Vessels in [7] aimed to decrease the time necessary to apply two particular WA's testing models. These models produce test paths to ensure good coverage of the WA. This approach succeeded to reduce the manual work of finding good test paths for a WA's.

The third approach presented in [9]. It proposed approach that significantly reduces the effort and cost of regression test suite generation. Because WAs evolve over time, they need to be continuously retested for change to ensure lasting components correctness and stability, thus assuring software quality. Hence, Regression testing needs to be carried out.

The final approach presented in [8]. It modeled the WAs into Finite State Machines (FSM) without using the source code and then applies the Genetic Algorithm for generating the test cases for testing. It deals with the problem of automatically testing of the WAs.

a Survey of WA testing was introduced in [11] which concluded the following: a FSM model approach was found to be the best test coverage approach. For test effectiveness, input validation approach, event flow approach, and system requirement model approach are on par with test-first design approach. It was also found that several test aspects of WA's functional testing namely interface testing and dynamic WA testing requires further contributions. Their overall evaluation result was that best testing results are when using test-first design approach and test case design approach are used.

Another Survey of WA testing was introduced in [10] which concluded that traditional application testing techniques can be enhanced and reused to test functional requirements of WAs, but non-functional requirements needs new and specific approaches to test them in WAs. Furthermore, it suggested defining appropriate TDD methods for WAs or use agile test process for such application.

The final survey we introduced in [12] where Kam & Dean concluded from their survey that there's no one single testing technique can be used to cover all bugs of a WA, and we need to integrate many techniques with each other to fully validate and verify a web application.

## 4. GUI Testing

A Graphical User Interface (GUI) testing is the process of testing software GUI to detect application errors and faults that restrict or change software required functionalities [13]. Desktop, mobile or web applications must have GUI, which is the visible part of the application that end users interact with. It was found that developing and testing the GUI takes up a significant part of the development effort, as much as 50–60 percent of the development time [14]. GUIs have basic characteristics that are different from traditional software, so traditional testing methodologies do not apply directly to GUI software [15, 16].

4.1 The Main characteristics of GUI Applications[15, 16]:

(1) Graphical orientation.
(2) Event-driven software and extremely large input space.
(3) Hierarchical structure.
(4) The variety of graphical objects, and the attributes of those objects.
(5) Hidden synchronization and dependencies.
(6) Object-oriented software programming.
(7) The wide range for user's interaction with GUI applications (as keyboard shortcut, a button click, a menu option)

Using conventional methodologies for GUI testing can result in increased time and expense[16].

Michael Turpin [5] provided the following guidelines to successful software testing suite: (1) GUIs require simulated user-generated events. (2) GUIs layout changes should not affect robust system, especially when using TDD. (3) Test coverage criteria, while it is important to ensure that all code is tested**.** Furthermore**,** He emphasized the importance of using regression testing to discover all remaining bugs of the system. Finally, test design and automatic testing approach must be conducted for this type of applications.

4.2 Some of GUI Testing Approaches

There are many authors who proposed techniques to perform automatic tests for GUI-based applications as [15, 17] that process events which automatically manipulate the GUI tests.

Reverse engineering technique was used in order to automatically generate GUI testing model [4, 18] . It was concluded that [18] if the application model is hard to be created manually, it is extremely hard to derive these models automatically. In this case manual exportation is used to improve automatic exploration of the model [4].

Genetic algorithm was used in [19] to automatically test GUI functionality. Graph models was driven to approximate all possible sequences of events that may be executed on the GUI, and then use graphs to generate test cases that achieve a specified coverage goal. It was concluded that feasible coverage of test suites can be increased, and the use of genetic algorithm outperforms the random algorithm. The main drawback of that approach is that the execution time increase which may prevent this approach from scaling to large applications mainly due to the increase of event dependencies that maybe generated.

A new technique called ALT was presented to generate GUI test cases in batches [20]. The problem was that the current fully automatic model-based test case generation tools are unable to discover the relationships between GUI events (e.g. one event enables the other event). Analysis of the run-time state of GUI widgets obtained from a previous test batch is used to obtain a new batch. Testing all possible event interactions that can lead to serious software failures were shown to be essential for successful testing. Consequently, software may need to be tested in all events interactions states.

A detailed survey was presented for existing GUI testing tools namely Abbot, Jacareto, Pounder, JFC and Marathon based on the events and fields needed [13], and Various GUI Testing approaches such as PTA, MBT, CIT and GAPs are studied with their event coverage. It was concluded that there's no single tool and approach fully covers all the testable items, and the performance of open source tool is yet to be studied with many applications.

We found that black-box testing of GUI-based Application could be achieved by executing sequences of events based on the GUI model, and researches have shown that testing a GUI-based Application from this perspective will find faults related not only to the GUI and its glue code, but to the underlying business logic of the application as well [17, 19, 21, 22].

## 5. Test-Driven Development

Test-Driven Development (TDD) is an approach of developing software where the test cases are developed that generally will not even compile, before it's related software code is implemented to pass those test cases [23, 24]. So there is no new functionality is developed unless all previous test cases run properly and its test case was developed [24]. TDD is one of the central techniques of Extreme Programming (XP). However, the impact of test-driven development on the business value of a project has not been studied so far [25]. With XP, developers do little or no up-front design before embarking in tight TDD cycles consisting of test case generation followed by code implementation [24].

IBM team [24] had a practice of using TDD in development and they concluded that the relatively inexperienced team realized about 50% reduction in Functional Verification Test defect density when compared with an experienced team who used an ad-hoc testing approach for a similar product. That indicates the importance of using TDD vs. traditional development and tests.

Developing GUI application using TDD is difficult [26, 27]. TDD requires all functionalities to be easily testable, before it can be implemented. So to implement GUIs by TDD, GUI code should be easily testable via automated unit tests and the test should be written before the user interface code. This can present a few difficulties since this will require the programmer to build a test for a GUI, which has not been seen [26].

In order to overcome these difficulties tools have been implemented to facilitate this process. For example a GUI testing tool called GTT [27], was implemented so test specifications can be developed independent of its application under test Thus, they can use TDD for GUI-based application.

An alternative to automated GUI system tests called Presenter First (PF) [28] was developed based on the Model-View-Presenter (MVP) design pattern. The PF gives test coverage corresponding directly to user stories. The PF is the behavior that corresponds directly to customer stories. The technique tends to delay working on the model until they uncover all system requirements.

## 6. Test Coverage

*"Coverage is the extent that a structure has been exercised as a percentage of the items being covered. If coverage is not 100%, then more tests may be designed to test those items that were missed and therefore, increase coverage"* [29].

Coverage criteria defined as set of rules to help determine whether a test suite has adequately tested a program, so it can guide testing process. Test Coverage helps in evaluating the effectiveness of software and it's a critical indicator of software quality. [15, 29]

Test coverage problems have an effect on GUI testing, and the automation may make things worse. It might be difficult to answer the questions "Does this set of test cases test everything that needs to be tested?" and "do some of these test cases overlap?", Especially when test cases are automatically generated by the automation suite. Due to the special nature of GUIs, normal code coverage criteria don't fit here, so one has to have new ones specified to examine test coverage in GUIs [26].

In paper [29] authors conclude that more work needs to be done in order to improve the current state of research in test coverage measurement and analysis.

## 7. Analysis of The State-of-the-art Testing Techniques of WA's

The importance of good quality WA's testing is increasing. Recently, many mission critical applications are designed as WA's, such as banking, or E-commerce systems where any failure may lead to irreparable damage of institutions.

Our conclusion from test approaches mentioned was:

1) There's no one single testing technique can be used to cover all bugs of a WA [12].
2) We can't directly use TDD approach in developing WAs, and we need to define appropriate TDD methods for WAs[10].
3) Because of the similarities between WA's and GUI Applications where both use GUI, hence tests done through their GUI, using algorithms such as genetic algorithm may be very successful to generate all available test paths for GUI but for a complete enterprise applications huge number of test cases must be tested, most of them have no impact and not important [19].
4) FSM model approach is the best for testing coverage [11].
5) Input validation approach, event flow approach, and system requirement model approach was found to be as good as the test-first design approach for test effectiveness [11].

From WA test approaches mentioned, we conclude that there is no one perfect approach to test WA. We need to integrate many techniques, or adapt some combinations of traditional Application testing techniques.

We have addressed some of GUI testing approaches because the GUI is the most similar Application to WAs, the nature of both of them has User Interface (UI) regardless UI type. Which we may find approach to adapt it on WA testing.

Recently many projects adopted TDD development because of its ability to discover as early as possible most of software failures. But TDD development has an issue dealing with applications that have GUI. Some approaches were shown to successfully handle this issue such as GTT and MVP.

Also we are concluded that perform GUI tests by executing sequences of events based on a model of the GUI will find faults related not only to the GUI and its glue code, but to the underlying business logic of the application as well. The ambiguity of what causes the faults the GUI or the business logic is one of the drawbacks of all of the current approaches. Testing the GUI through business logic, will not guarantee that all the combinations of events that may cause faults in the GUI are tested. In the other hand, using the GUI to test all possible application logic is very difficult and some approaches will not cover all the functionalities, and some approaches will generates redundant irrelevant tests among those that need to be tested which make it practically impossible to be used.

So a new testing technique for WA's need to be discovered in order to practically perform tests have the right testing coverage practically and permit developing techniques such as TDD.

## 8. Specifications of a Novel and Improved WA's Testing Technique

From the analysis of the current state-of-the-art testing techniques the characteristics of the new approach has emerged. This approach will be based mainly on a novel Gray-Box testing technique where a three sets of tests need to be generated from directly system requirements. The first set of tests needs to be directed to the functionalities of the GUI independent of the business logic. Other set of tests should be directed to the application logic without any use of the GUI and finally a set of tests needed for the screen transitions. That will guarantee that the number of tests will be as small as possible, nevertheless, its testing coverage is complete for all possible faults of the system. It will be required to prove that if S is the complete set of tests with total coverage of all testable functionalities. Hence, $M \subseteq S$, where M is the minimal set of total coverage of the application. If P is the set of generated tests from the system requirements, and $P = p1 \cup p2 \cup p3$ where p1 is the set of application logic functionalists tests, p2 is the set of GUI functionalities tests and P3 is the set of all screen transitions tests then

$$S \subseteq f (p1 \times p2 \times p3)$$

Where $f$ is a function that take the Cartesian product of p1, p2 and p3 and produce a set where each of its elements is f(<t1, t2, t3>) where t1 ∈ p1, t2 ∈ p2 and t3 ∈ p3. The function f integrates 3 tests into one test. Since $M \subseteq S \rightarrow M \subseteq f (p1 \times p2 \times p3)$. Therefore, $f$ (p1 X p2 X p3) contain tests for total tests coverage. Note that $f$ (p1 X p2 X p3) is not minimal set therefor it contain a lot of irrelevant tests, nevertheless, the tests that need to be examined are P1, P2 and P3 independently. The set $f$ (p1 X p2 X p3) is an implied never materialized. Consequently, the number of tests needed to be run is much less that the cardinality of $f$ (p1 X p2 X p3) and less than the cardinality of M.

## 9. Conclusion

Web Applications (WA's) failures may lead to collapse of the institutions, therefore the importance for good quality WA's increasing over the time.

This research analyzed the main testing approaches for WA's and GUI applications. Also we have discussed Test-Driven Development approach and its effects on the current development.

We found some issues related to test approaches of WA's that require using more than one approach that covers all of WA's testing criteria. Because of the similarities between WA's and GUI Applications where both use GUI, We have discussed GUI testing approaches and its impact that may give us the change to adapt one approach for using it to WA's testing.

Finally we proposed a novel approach for WA testing. This approach aims to solve testing issues which are found in current approaches of WAs.

## References

[1] C. Eaton and A. M. Memon, "Chapter 5 Advances in Web Testing," in *Advances in Computers*. vol. 75, V. Z. Marvin, Ed., ed: Elsevier, 2009, pp. 281-306.

[2] L. Luo, "Software Testing Techniques," *Carnegie Mellon University,Pittsburgh, USA*.

[3] A. Jangra, G. Singh, J. Singh, and R. Verma, "Exploring testing strategies," *International Journal of Information Technology and Knowledge Management,* vol. 4, pp. 297-299, 2011.

[4] A. C. R. Paiva, J. C. P. Faria, and P. M. C. Mendes, "Reverse engineered formal models for GUI testing," presented at the Proceedings of the 12th international conference on Formal methods for industrial critical systems, Berlin, Germany, 2008.

[5] M. T. urpin, "Survey of GUI Testing Processes," *UMM Computer Science Seminar II* 2008.

[6] F. Belli and M. Linschulte, "On "Negative" Tests of Web Applications," *Annals of Mathematics, Computing & Teleinformatics,* vol. 1, No. 5, pp. 44-56, 2007.

[7] S. Vessels, "Automatic Generation of Artifacts for Two Web Application Testing Models " 2011.

[8] K. Singh, R. Kumar, and I. Kaur, "Testing Web Based Applications Using Finite State Machines Employing Genetic Algorithm," *International Journal of Engineering Science and Technology (IJEST),* vol. 2(12), 2010.

[9] Gagandeep and J. Sengupta, "Automatic Generation of Regression Test Cases for Web Components using Domain Analysis and Modeling," *International Journal of Computer Applications* vol. 11, 2010.

[10] G. A. D. Lucca and A. R. Fasolino, "Testing Web-based applications: The state of the art and future trends," *Inf. Softw. Technol.,* vol. 48, pp. 1172-1186, 2006.

[11] M. Y. Suhaila, W. K. W. M. Nasir, Member, and IAENG, "An Outlook of State-of-the-Art Approaches in Functional Testing of Web Application " *International MultiConference of Engineers and Computer Scientists,* vol. 1, 2011.

[12] B. Kam and T. R. Dean, "Lessons Learned from a Survey of Web Applications Testing," presented at the Proceedings of the 2009 Sixth International Conference on Information Technology: New Generations, 2009.

[13] J. Prabhu and N. Malmurugan, "A Survey on Automated GUI Testing Procedures " *European Journal of Scientific Research* vol. 64, pp. 456-462, 2011.

[14] A. M. Memon, "A comprehensive framework for testing graphical user interfaces," *Ph.D. dissertation, University of Pittsburgh,* 2001.

[15] A. M. Memon, M. L. Soffa, and M. E. Pollack, "Coverage criteria for GUI testing," *SIGSOFT Softw. Eng. Notes,* vol. 26, pp. 256-267, 2001.

[16] P. Li, T. Huynh, M. Reformat, and J. Miller, "A practical approach to testing GUI systems," *Empirical Softw. Engg.,* vol. 12, pp. 331-357, 2007.

[17] Q. Xie and A. M. Memon, "Using a pilot study to derive a GUI model for automated testing," *ACM Trans. Softw. Eng. Methodol.,* vol. 18, pp. 1-35, 2008.

[18] A. Memon, I. Banerjee, and A. Nagarajan, "GUI Ripping: Reverse Engineering of Graphical User Interfaces for Testing," presented at the Proceedings of the 10th Working Conference on Reverse Engineering, 2003.

[19] S. Huang, M. B. Cohen, and A. M. Memon, "Repairing GUI Test Suites Using a Genetic Algorithm," presented at the Proceedings of the 2010 Third International Conference on Software Testing, Verification and Validation, 2010.

[20] X. Yuan and A. M. Memon, "Iterative execution-feedback model-directed GUI testing," *Information and Software Technology,* vol. 52, pp. 559-575, 2010.

[21] T. Tuglular, C. A. Muftuoglu, O. Kaya, F. Belli, and M. Linschulte, "GUI-Based Testing of Boundary Overflow Vulnerability," presented at the Proceedings of the 2009 33rd Annual IEEE International Computer Software and Applications Conference, 2009.

[22] P. A. Brooks, B. P. Robinson, and A. M. Memon, "An Initial Characterization of Industrial Graphical User Interface Systems," presented at the Proceedings of the 2009 International Conference on Software Testing Verification and Validation, 2009.

[23] J. Rantanen, "Acceptance Test-Driven Development with Keyword-Driven Test Automation Framework in an Agile Software Project," Department of Computer Science and Engineering, Helsinki University Of Technology, 2007.

[24] E. M. Maximilien and L. Williams, "Assessing test-driven development at IBM," presented at the Proceedings of the 25th International Conference on Software Engineering, Portland, Oregon, 2003.

[25] M. M. Müller and F. Padberg, "About the Return on Investment of Test-Driven Development " *International Workshop on Economics-Driven Software Engineering Research EDSER-5,* 2003.

[26] M. Mikkolainen, "Automated Graphical User Interface Testing," www.cs.helsinki.fi/u/paakki/mikkolainen.pdf, 2006.

[27] W.-K. CHEN, Z.-W. SHEN, and T.-H. TSAI, "Integration of Specification-based and CR-based Approaches for GUI Testing " *Information Science and Engineering,* vol. 24, pp. 1293-1307, 2008.

[28] M. Alles, D. Crosby, C. Erickson, B. Harleton, M. Marsiglia, G. Pattison, and C. Stienstra, "Presenter First: Organizing Complex GUI Applications for Test-Driven Development," presented at the Proceedings of the conference on AGILE 2006, 2006.

[29] M. Shahid, S. Ibrahim, and M. N. r. Mahrin, "A Study on Test Coverage in Software Testing " *International Conference on Telecommunication Technology and Applications,* vol. 5, 2011.

**Mostafa Kandil**: received the B.Sc. degree from the Department of Information Systems, Faculty of Computers and Information, Cairo University in 2007. He is currently working toward the M.Sc. degree from the same department, and also Senior Software engineer at the Center of Development of Computers and Information Systems

**Ehab Hassanein:** Currently Lecturer at the Department of Information Systems at the Faculty of Computers and Information, Cairo University. In addition, serving as the vise-manger of the center of development of computer and information Systems. Between 2000-2006 worked at Lucent Technologies, Member of the Technical Staff in Mobility (UMTS)at Naperville, IL. USA. Form 1995-1997 was Assistant Professor in the Computer Science Department of The Knowledge Systems Institute Skokie, IL. USA. From 1993-1995 Software Development Consultant at HAQ, Inc. Des Plaines, IL. USA. Received his PhD from the department of Electrical Engineering and Computer Science Northwestern University Evanston, IL.

**Sherif Mazen:** currently Associate Professor of Information Systems and President of the Internal Auditing committee for Quality Assurance accreditation at Faculty of Computers & Information, Cairo University (URL: www.fci.cu.edu.eg). Dr. Mazen research areas: IS Strategic Planning, IS Analysis & Design, IS Project Management, IS Auditing, IS Quality Assurance, E-commerce...etc. Also, He holds a Ph.D. - in December 1982 - in "Applied Informatics in Hospital Management" from Toulouse (I) University, France. He teaches several graduate and post-graduate courses like: Impact of Information Technologies on Management, Systems Analysis & Design, IS Project Management, Information Centers Management, IS auditing …etc.